# To Share or Not to Share: Orchestrating Trustworthy Data in Global Value Chains

Han-Teng Liao, *GBS Today, Member, IEEE*, and Chang-Yi Kao, *Soochow University, Taipei, Taiwan*, *Member, IEEE*.

***Abstract*—** As the EU Carbon Border Adjustment Mechanism (CBAM) approaches, the global semiconductor value chain faces growing structural tensions between regulatory transparency and data sovereignty. This article proposes a RegTech reference architecture using the International Data Spaces (IDSA) framework to orchestrate trustworthy environmental telemetry across the semiconductor-petrochemical nexus. The framework distinguishes the mandatory CBAM requirements from voluntary Science Based Targets initiative (SBTi) frameworks, while addressing the additive complexities of the Safe-and-Sustainable-by-Design (SSbD) framework. Moving beyond standard linear technology stacks, we introduce a prospective roadmapping methodology that transforms upstream physical vulnerabilities into circular, negative feedback loops. Focusing on the Taipei and Penang technology corridor, the article details how sovereign data exchange enables Digital Product Passports (DPPs) to drive Global Business Services (GBSs) capability demands. Finally, we discuss the integration of Agentic AI for autonomous compliance and FinTech green financing, providing a scalable blueprint for global industrial clusters to achieve sovereign, sustainable, and transparent value chains.

## I. The Strategic Pole Vault: Beyond Voluntary Pledges

FOR technology managers in the Asia–Pacific region, the transition to the 2026 regulatory cycle is not a trap but a test of elevation — a pole vault moment. Success depends on building the capabilities to clear new regulatory bars and secure continued access to European markets. A persistent misconception remains: that high-level adherence to the **Science Based Targets initiative (SBTi)** is sufficient to meet the **Carbon Border Adjustment Mechanism (CBAM)** requirements. This is a high-stakes error; managers must distinguish voluntary vs. mandatory frameworks:

- **SBTi (Science Based Targets initiative):**
  A voluntary, corporate-level disclosure framework designed to **demonstrate long-term commitments** to net-zero pathways and aggregate emissions reductions, primarily oriented toward investor relations, brand positioning, and word-of-mouth, but not legally binding.
- **CBAM (Carbon Border Adjustment Mechanism):**
  A mandatory EU trade regulation requiring **product-level verification of actual embedded emissions** for goods entering the Union that demands granular, "location-based" values tied to specific production lines, facilities, and supply chain nodes, enforced through customs procedures and trade penalties.

While SBTi helps firms **communicate ambition**, CBAM **enforces accountability**. Managers who conflate the two risk regulatory non-compliance, market exclusion, and reputational damage.

### *A. Implications: Massive Data Workflows for Accountability*

For semiconductor giants and their upstream petrochemical providers, 2026 marks the end of reliance on "default values" and the beginning of mandatory **data transparency** [1]. The first Taiwan carbon fee cycle in 2026 collects total NT$4.97 billion (US$158.08 million), with 123 semiconductor facilities contributing about NT$2.2 billion, larger than the electricity supply (NT$635 million), steel (NT$400 million), and concrete (NT$130 million) combined. This clearly signals that key Asia-Pacific players such as Taiwan are ready to embrace the transition from voluntary pledges to auditable outcomes. It also suggests that the semiconductor sector and the AI sector driving its demand must take the lead to clean itself.

While the initial CBAM targets foundational sectors, impending scope expansions mean CBAM is poised to require firms to disclose actual emissions embedded in products such as wafers, solvents, and specialty chemicals soon. This shift forces managers to move beyond sector-wide baselines and toward **line-by-line measurement and reporting**. Investing in CBAM data infrastructure and workflows now will compound process and tacit knowledge.

For more complex regulatory demands beyond carbon emissions, the EU **Safe-and-Sustainable-by-Design (SSbD)** framework introduces additional obligations for chemical safety and lifecycle transparency [2], [3], [4], requiring firms to demonstrate that critical inputs comply with EU sustainability criteria. At the same time, International Telecommunication Union (ITU) digital innovation toolkits and standards — global

Han-Teng Liao (0000-0003-1081-5599) is an independent researcher with an IEEE TEMS membership, based in Penang and Taipei. He was with the Oxford Internet Institute, UK, the United Nations University, a university in China and Oxford Roadmapping Technology Services (e-mail: h.liao@ieee.org).

Chang-Yi Kao (0000-0003-0075-0787) is with the Department of Computer Science and Information Management, Soochow University, Taipei, Taiwan (e-mail: edenkao@scu.edu.tw).

Color versions of one or more of the figures in this article are available online at http://ieeexplore.ieee.org

technical standards for telecommunications and ICT interoperability — provide ICT-sector decarbonization references for regulators and industry consortia [5], [6]. For managers, this means compliance is not only about emissions reporting but also about aligning with global standards for data sovereignty (the principle that organizations retain control over how their data is used and shared) and secure information flows. These workflows are likely to become the future of regulatory compliance using Regulatory Technology (RegTech), a field that uses digital tools to automate compliance.

International Data Spaces Association (IDSA) data sovereignty standards provide an information system architecture for securing data sovereignty across corporate and industrial ecosystems, allowing trusted partners to access under terms of use [7]. It has been discussed for the Internet of production in the context of the European GDPR guideline [8] and the design of sustainable business models for European manufacturing [9]. The stakeholder analysis of real-world use cases show that such modular and collaborative models create strategic and economic values for stakeholders, confirming the usefulness of data sovereignty, trust, and interoperability as design principles [9]. Indeed, recent work has validated the feasibility of cross-data space interworking for $CO_2$ footprint monitoring between Europe's IDS and Japan's CADDE frameworks, demonstrating that sovereign data exchange can support carbon accountability across regions [10], as the critical pilot between the European and Asia–Pacific regions.

### *B. Question: Compliance Innovations in Global Value Chains*

A critical challenge remains: How might we design IDSA-based RegTech to transform compliance burdens (e.g., CBAM and SSbD) into growth points for sustainable, multi-stakeholder business models across wider Global Value Chains (GVC)? Particularly for technology managers and industry strategists, how might we integrate such data-driven accountability workflows for the high-growth AI and semiconductor sectors, which drive both innovations and emissions at the same time?

## II. Making Nvidia Five-Layer AI Cake Circular

Nvidia CEO Jensen Huang, in a 2026 keynote, framed AI as essential infrastructure, "*like electricity and the internet*", unfolding as a five-layer industrial stack — **energy → chips → infrastructure → models → applications**. Each layer reinforces the other and pulls demand downward to the physical foundations of computation [11].

Huang's metaphor of the "AI cake" emphasizes exponential growth in **tokens** and their downstream economic impacts. The rising energy demands of this paradigm can already be observed in the energy paradox of Taiwan's sovereign AI development [12], and Nvidia's explicit lobby for expanded electricity infrastructure within semiconductor hubs such as Taiwan [13].

### *A. Method: Prospective Roadmapping*

To answer our core research question, this article utilizes a prospective roadmapping methodology anchored in a socio-technical systems approach [14]. Unlike retrospective tracking, prospective roadmapping systematically projects future regulatory and industrial milestones to engineer technology infrastructures ahead of market enforcement. This method translates high-level policy constraints into concrete, actionable capabilities across two distinct design phases.

This article builds upon the detailed analysis of the International Roadmap for Devices and Systems (IRDS) environmental documents [15], demonstrated in our forthcoming work [16], [17]. Corresponding to the physical layers of the AI cake—specifically Energy, Chips, and Infrastructure—impending EU-related compliance, including CBAM and SSbD, is synthesized as negative feedback loops to make the five-layer cake circular. By cross-referencing these upstream vulnerabilities as critical signals, the resulting reference architecture provides engineering managers with an active, forward-looking blueprint to transform automated compliance from a lagging operational cost into a long-term competitive capability.

### *B. The Missing Negative Feedback Loop*

While Huang's vision captures economic and technological momentum, it overlooks the **environmental impacts** (as indicated by [15]). **Negative feedback loops** can monitor and control those impacts. AI and cloud computing companies often position themselves as leaders in ICT-sector decarbonization. Yet their emissions are **outsourced upstream** — to the **semiconductor–petrochemical nexus** that supplies the chips, solvents, and materials enabling AI factories. This outsourcing creates a managerial blind spot rapidly becoming the focus of global regulatory scrutiny.

Fig. 1 shows how such feedback loops stabilize the AI stack. The exponential growth of tokens may generate economic value on the left-hand "*Electrons to Tokens*" stack, but, without integrated accountability mechanisms, environmental costs accumulate invisibly in upstream production on the right-hand "*Atoms to Accountability*" circuit. CBAM and SSbD requirements expose these hidden costs, introducing negative feedback loops to manage impacts.

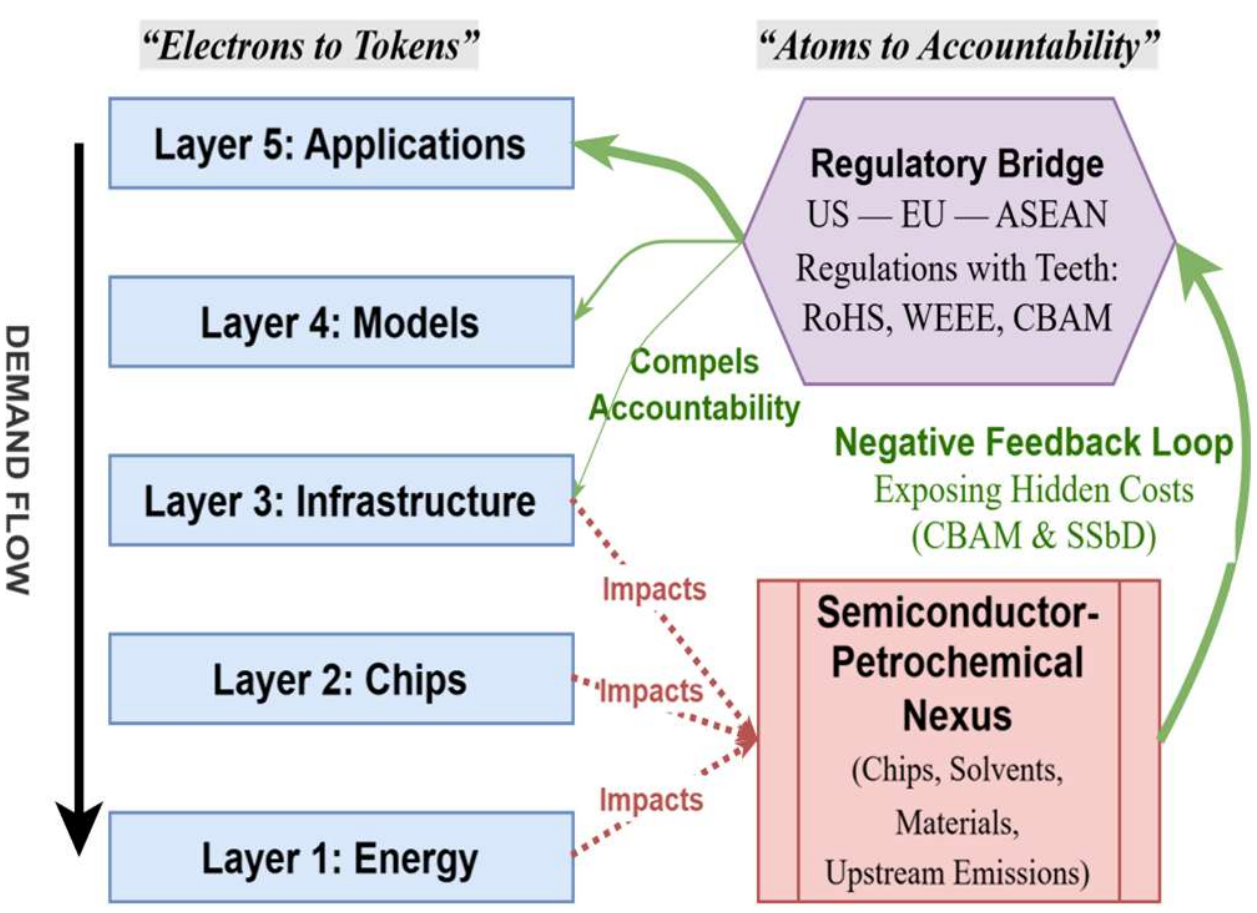

Fig. 1. The "Atoms-to-Accountability" Socio-Technical Governance Loop for Cross-Border Compute Infrastructure (Adapted and expanded from [16])

Given the differences between the United States and the European Union on digital services and trade, and ASEAN's ongoing push for digital economy frameworks and paperless trade standards, technology managers and systems engineers must prepare to bridge regulatory gaps proactively before geopolitical demands arrive.

As indicated in the top-right of Fig. 1, the proven enforcement strength of EU instruments such as **CBAM, RoHS (Restriction of Hazardous Substances), and WEEE (Waste Electrical and Electronic Equipment Directive)** demonstrates that global markets respond to regulations "with teeth." Consequently, the industrial AI stack can be regulated by addressing planetary limitations through a circular "Atoms to Accountability" perspective.

## III. Architecting Data Sharing Solutions

If technology managers do not lead in making AI environmentally accountable, regulators will — and at greater financial and environmental cost. It is therefore critical to **architect data and Agentic AI infrastructures** that support the latest **RegTech** and **FinTech** innovations, especially in CBAM and SSbD compliance. Managers who treat accountability as a design principle, rather than a reporting obligation, will be better positioned to clear the regulatory bar and convert compliance into competitive advantage.

### *A. Why Data Sharing Comes First*

AI and cloud computing companies claim environmental responsibility, yet outsource emissions and environmental impacts upstream to the semiconductor–petrochemical nexus, creating systematic opacity in accountability. Data centers may appear efficient on paper, but the solvents, precursors, and hydrocarbon inputs enabling chip production and running remain hidden from accountability.

Without upstream accountability and downstream feedback loops, AI compute risks consuming natural resources regardless of negative impacts. Managers must prioritize data sharing architectures that integrate upstream and downstream material flows into unified compliance workflows. Sovereign data spaces enable secure data exchange channels, connecting petrochemical suppliers to chip manufacturers to electronics companies in auditable, verifiable lifecycles.

To operationalize this architecture across multi-stakeholder ecosystems, compliance metrics must be anchored to a Digital Product Passport (DPP)—a standardized, machine-readable record that tracks a product's embedded emissions, chemical safety profiles, and circularity data as it traverses global value chains. This unified approach simplifies regulatory proof and enables real-time traceability. As illustrated in Fig. 2, the DPP serves as the primary data vehicle, utilizing secure exchange infrastructures to bridge upstream complexity with downstream operational workflows. Such **continuous, product-level data flows** now support the CBAM, SSbD, RoHS and WEEE compliance workflows, where **Agentic AI** and **data sovereignty architectures** become indispensable enablers.

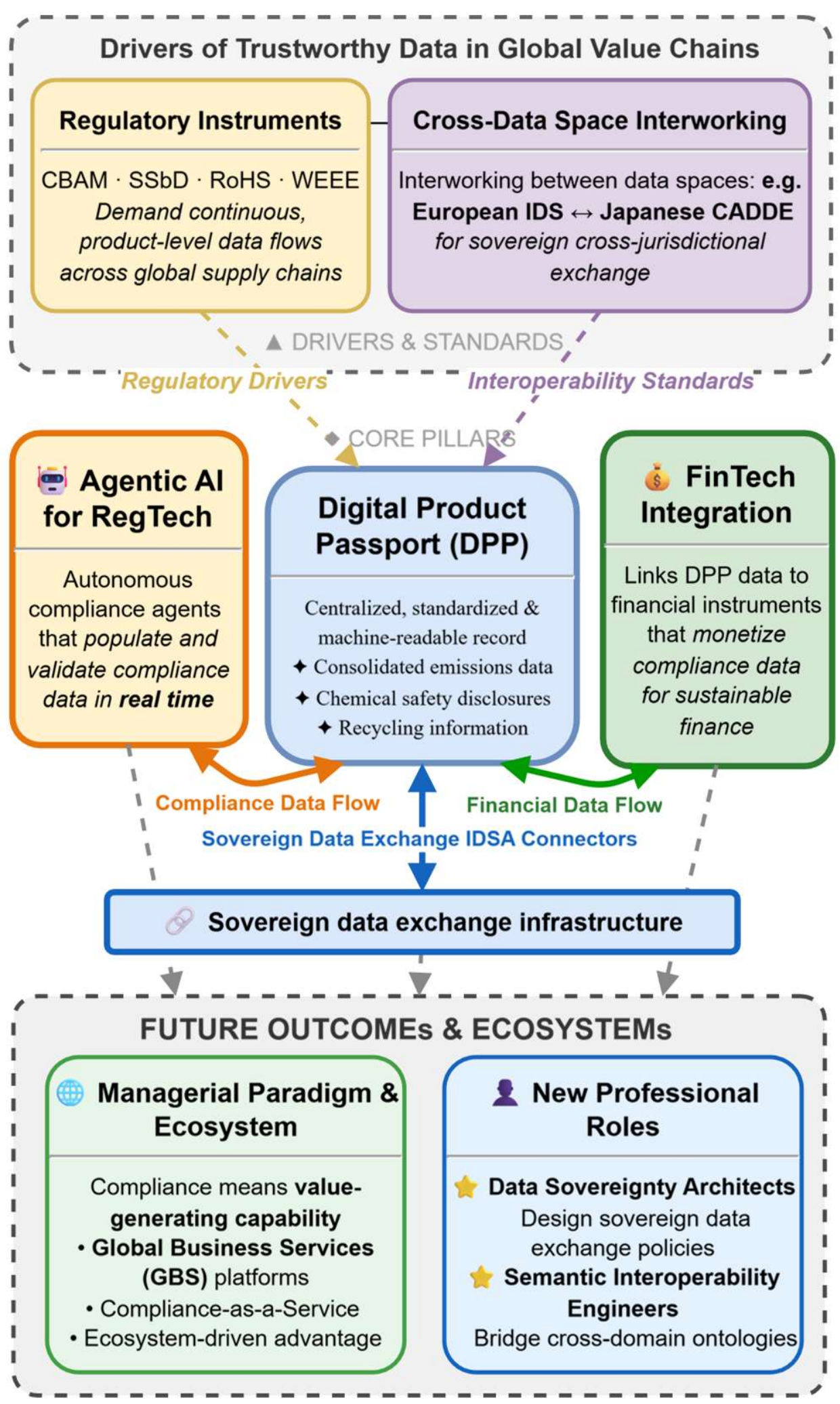

Fig. 2. Conceptual RegTech Reference Architecture: Utilizing sovereign data space infrastructures to route Digital Product Passports (DPP) across cross-border value chains, integrating upstream environmental telemetry (Sections III.B and C) with downstream automated agentic workflows (Section III.D)

### *B. Upstream Complexity: Carbon Emissions*

The semiconductor industry does not begin at the wafer fab, but in petroleum refining and chemical synthesis. To satisfy CBAM, managers must trace upstream to precursors: high-purity solvents such as **isopropyl alcohol (IPA)** and other energy-intensive materials targeted for carbon pricing. For instance, Taiwan's LCY Chemical and Formosa Plastics manufacture the critical solvents for wafer cleaning, while Malaysia's PETRONAS—via its acquisition of Perstorp—controls segments directly bound by EU regulations. Rather than serving as passive suppliers, these firms represent cross-border compliance opportunities.

As conceptualized in the macro-governance loop (Fig. 1), managing these upstream nodes requires verifying actual emissions data. The reference architecture (Fig. 2) operationalizes this by routing sovereign data exchanges to transform corporate pledges into auditable outcomes. The Taipei–Penang technology corridor could construct and validate these data channels now, establishing regulatory sandboxes, to manage CBAM and SSbD requirements before punitive tariffs are enforced at EU borders.

*C. Upstream Complexity: Harmful Processing*

SSbD adds a second compliance layer, ensuring that chemicals are non-toxic and environmentally benign during production and use. However, managers often fall into a **double-reporting trap**, building isolated teams for carbon (CBAM) and chemical safety (SSbD). This siloed approach is inefficient and duplicative. By investing in IDSA-based architectures, firms can realize a **"collect once, report twice"** paradigm, strengthening data integrity, and accelerating compliance timelines.

The ICT sector has decades of experience adapting to regulations such as RoHS and WEEE, which integrated compliance into product design, supply chain management, and end-of-life recovery processes. Firms that integrated these requirements early gained measurable competitive advantage. This historical precedent demonstrates that **early investments in compliance pay lasting dividends**.

*D. Data Sovereignty and Agentic Workflows*

With the structural data channels mapped onto Fig. 2, three automated system enablers transform the DPP from a passive logging entry into an active risk-mitigation tool:

- **Agentic AI for RegTech:** Autonomous agents populate and validate DPPs in real time, ensuring continuous adherence to CBAM and SSbD.
- **FinTech Integration:** Linking DPP metadata to financial instruments — carbon pricing, green bonds, or sustainability-linked loans — transforms compliance into financial leverage for decarbonization investment.
- **IDSA Connectors:** Interoperable infrastructures ensure that DPP operations are seamless across jurisdictions, allowing firms to "collect once, report twice" without exposing proprietary secrets.

The cross-border feasibility of this model is validated by the Europe-Japan (IDS-CADDE) data sharing pilot [10]. This convergence redefines compliance from an isolated administrative cost center into an active, **value-generating capability**. Firms leveraging Physical and Agentic AI to automate CBAM, SSbD, RoHS, and WEEE reporting reduce regulatory risk while unlocking new financing channels tied to sustainability performance.

## IV. Managerial Implications and Future Practice

The transition from voluntary sustainability reporting to enforceable regulatory mandates requires technology managers to re-engineer value chain workflows across global operations.

The pole vault metaphor underscores that managers face a rising bar rather than a cliff. Clearing this bar requires preparation, momentum, and the right tools. With the arrival of Agentic AI and physical AI —where autonomous software agents orchestrate complex data workflows while edge-intelligent hardware monitors real-world physical processes and enforces compliance in real time—and the maturation of IDSA ecosystems, managers can begin designing RegTech solutions that automate compliance, orchestrate trustworthy data exchange, and convert regulatory pressures into competitive advantage.

This transition reframes compliance as capability-building. Those who invest early in data workflows, interoperability standards, and sovereign exchange architectures will not only meet CBAM and SSbD requirements but also position themselves as leaders in the emerging RegTech- or FinTech-enabled Global Business Services (GBS) and Global Capability Center (GCC) sectors [18], [19]. Moreover, ICT sectors—including telecom and cloud computing giants—must extend their role beyond technical enablement. They must work with semiconductor and petrochemical counterparts into a deep audit regime, covering carbon emissions (especially Scope III) and full life-cycle compliance with SSbD procurement, tracking, and reporting, thereby integrating RegTech, FinTech and Industry 5.0 ecosystems.

*A. Key Global Value Chains: Semiconductor-Petrochemical*

Looking ahead, the orchestration of trustworthy data in semiconductor–petrochemical value chains will require not only technical infrastructures but also organizational innovation [20]. The evolution of GBS and GCC offers promising service-platform ecosystems for embedding digital compliance capabilities into shared service models. Future research should examine how they can incubate new professional roles — including **Data Sovereignty Architects** and **Semantic Interoperability Engineers** — to embed continuous, product-level accountability into global supply chains and accelerate digital innovation practices. In practice, integrating **digital product passports (DPPs)**, **Agentic AI compliance modules**, and **sovereign data exchange standards** into GBS and GCC business models could accelerate the transition from regulatory burden to digital innovation, positioning firms at the forefront of Industry 5.0.

*B. Strategic Stakeholder Roadmap*

To operationalize these modular value-capture models within the complex semiconductor-petrochemical nexus, technology managers must explicitly evaluate strategic imperatives, operational risks, and success benchmarks across distinct organizational scales. Compliance with CBAM, SSbD, RoHS, and WEEE becomes a capability opportunity to leverage both Physical and Agentic AI. To gain compliance automation advantages, Table I summarizes a cross-layer alignment framework spanning four stakeholder roles — from regulators and enterprises to upstream suppliers and RegTech/FinTech startups and small- and medium-sized enterprises (SMEs).

Table I
STRATEGIC ALIGNMENT MATRIX FOR GVC COMPLIANCE ORCHESTRATION

| **Ecosystem Role** | 🌍 **Strategic Imperatives** | 👷 **Key Risks and Trade-offs** | 📏 **Benchmark Examples** |
|---|---|---|---|
| 🏛 Regulators and Standards Bodies (e.g., EU, ITU-T, Customs) | **Mandate Standardization:** Enforce predictable, legally binding compliance baselines (**CBAM, SSbD, RoHS, WEEE**) while recognizing sovereign data exchange protocols. | **Fragmented Jurisdictions:** High geopolitical and technical friction when reconciling disparate regional standards (e.g., bridging EU IDS with Asian data tracking architectures). | **Benchmark:** Elimination of administrative duplication by transitioning to universally accepted, interoperable Digital Product Passport (DPP) ontologies. |
| 🌐 Semiconductor Enterprises (e.g., TSMC, Infineon, ESMC) | **Value Chain Orchestration:** Lead regional IDSA hubs to establish uniform semantic ontologies across multi-tier supplier networks. | **Legacy Integration:** High upfront capital and time required to retrofit automated IDSA connectors into deeply siloed, legacy ERP systems. | **Benchmark:** 100% real-time, verified embedded emissions tracking across all Tier-1 chemical and material suppliers within 18 months. |
| 🏭 Petrochemical Upstream Suppliers (e.g., PETRONAS, LCY Chemical, Formosa Plastics | **Value Chain Alignment and Data Sovereignty:** Deploy IDS Connectors to protect data while safely provisioning mandatory one. | **Resource Constraints:** Navigating "double-reporting traps" due to limited specialized compliance capability. | **Benchmark:** Transition from default carbon values to verified, line-by-line actual data profiles, including SSbD, using DPPs. |
| 📑 **RegTech and FinTech Startups and Providers, including SMEs** *(e.g., IntegrityNext)* | **Capability-as-a-Service:** Build agile, interoperable AI agents capable of plugging into regional data spaces to automate compliance tasks. | **Interoperability Issues:** High data frictions when navigating cross-jurisdictional standards, and shifting regulatory requirements. | **Benchmark:** Reduction in time-to-compliance certificate issuance; verified ingestion of compliance data by green finance auditors. |

In sum, managers who treat compliance as a pole vault — a chance to elevate capabilities rather than avoid cliffs — will be best positioned to lead semiconductor value chains into trusted, sustainable, and globally competitive futures. Major Asian and European manufacturers should begin investments and standardization efforts.

## ACKNOWLEDGMENT

The authors gratefully acknowledge the editors and anonymous reviewers of the *IEEE Engineering Management Review* for their guidance in elevating this manuscript. We also express our gratitude for the technical standards, roadmap parameters, and frameworks provided by the International Telecommunication Union (ITU), the IEEE IRDS, and the International Data Spaces Association (IDSA), which were instrumental in shaping the ecosystem innovations, sustainability, and data sovereignty concepts presented in this work.

## REFERENCES

[1] Focus Taiwan, "Taiwan collects nearly NT$5 billion in first carbon fee cycle," *CNA English News*, The Central News Agency (CNA), Jun. 03, 2026. Accessed: Jun. 08, 2026. [Online]. Available: https://focustaiwan.tw/business/202606030008

[2] K. Schwirn *et al.*, "The European Commission's safe and sustainable by design framework: bridging innovation and legislation," *Environ. Sci Eur.*, vol. 37, no. 1, p. 189, Nov. 2025, doi: 10.1186/s12302-025-01246-y.

[3] P. Fantke and P. Mankong, "Insights from life cycle assessment to inform chemical substitution, alternatives assessment and safe and sustainable-by-design," *Environ. Sci. Technol.*, vol. 60, no. 2, pp. 1596–1611, Jan. 2026, doi: 10.1021/acs.est.5c12521.

[4] C. Apel *et al.*, "Safe and Sustainable by Design (SSbD) landscape: Mapping state of the art approaches along the innovation process and developing good practices," *Sustain. Chem. Pharm.*, vol. 50, p. 102355, Apr. 2026, doi: 10.1016/j.scp.2026.102355.

[5] A. Serpe *et al.*, "2002-2022: 20 years of e-waste regulation in the European Union and the worldwide trends in legislation and innovation technologies for a circular economy," *RSC Sustain.*, vol. 3, no. 3, pp. 1039–1083, Mar. 2025, doi: 10.1039/d4su00548a.

[6] H.-T. Liao, C.-L. Pan, and Y. Zhang, "Smart digital platforms for carbon neutral management and services: Business models based on ITU standards for green digital transformation," *Front. Ecol. Evol.*, vol. 11, p. 1134381, Mar. 2023, doi: 10.3389/fevo.2023.1134381.

[7] B. Otto, M. ten Hompel, and S. Wrobel, "International Data Spaces," in *Digital Transformation*, R. Neugebauer, Ed., Berlin, Heidelberg: Springer, 2019, pp. 109–128. doi: 10.1007/978-3-662-58134-6_8.

[8] M. Jarke, "Data Sovereignty and the Internet of Production," in *Advanced Information Systems Engineering*, S. Dustdar, E. Yu, C. Salinesi, D. Rieu, and V. Pant, Eds., Cham: Springer International Publishing, 2020, pp. 549–558. doi: 10.1007/978-3-030-49435-3_34.

[9] E. Politi *et al.*, "Designing Sustainable Business Models for Data Spaces: Insights from the Industrial Sector," *IEEE Access*, pp. 1–1, 2026, doi: 10.1109/ACCESS.2026.3695646.

[10] J. R. Santana *et al.*, "On the Need of International Cross-Data Space Interworking: An EU–Japan Case Study," *IT Prof.*, vol. 27, no. 5, pp. 59–65, Sep. 2025, doi: 10.1109/MITP.2025.3545214.

[11] J. Huang, "AI is a 5-layer cake," NVIDIA Blog. Accessed: Apr. 18, 2026. [Online]. Available: https://blogs.nvidia.com/blog/ai-5-layer-cake/

[12] S. Huang, "The energy paradox of Taiwan's sovereign AI ambition," *Center for Asia-Pacific Resilience and Innovation (CAPRI) Analysis*, Mar. 2026. [Online]. Available: https://caprifoundation.org/the-energy-paradox-of-taiwans-sovereign-ai-ambition/

[13] L. Wang, "Nvidia CEO touts surge in spending in Taiwan," *The Taipei Times*, Taipei, p. 1, May 28, 2026.

[14] A. Tuominen and T. Ahlqvist, "Is the transport system becoming ubiquitous? Socio-technical roadmapping as a tool for integrating the development of transport policies and intelligent transport systems and services in Finland," *Technol. Forecast. Soc. Chang.*, vol. 77, no. 1, pp. 120–134, Jan. 2010, doi: 10.1016/j.techfore.2009.06.001.

[15] IEEE, "Environment, safety, health & sustainability (ESHS): environmental sustainability of the semiconductor facilities (ESSF)," International Roadmap for Devices and Systems (IRDS), 2024. [Online]. Available: https://irds.ieee.org/images/files/pdf/2024/2024IRDS_ESHS-ESSF.pdf

[16] H.-T. Liao and K. Ang, "From stacks to circuits: a regenerative socio-technical roadmap for AI infrastructure within planetary boundaries," in *2026 IEEE International Conference on Engineering, Technology, and Innovation (ICE/ITMC)*, forthcoming 2026.

[17] K. Ang and H.-T. Liao, "Scoping review of AI, metrology, and ESG in the semiconductor sector: implications for Safe and Sustainable by Design (SSbD)," in *2026 IEEE International Conference on Engineering, Technology, and Innovation (ICE/ITMC)*, forthcoming 2026.

[18] H.-T. Liao and K. Ang, "Orchestrating the twin transition in multinational corporations: technology roadmapping for green and digital global business services," in *2026 IEEE International Conference on Engineering, Technology, and Innovation (ICE/ITMC)*, forthcoming 2026.

[19] S. K. Jha and A. Seth, "Global capability centres: emerging opportunities and challenges," *IIMB Manag. Rev.*, vol. 37, no. 3, p. 100595, Sep. 2025, doi: 10.1016/j.iimb.2025.100595.

[20] H.-T. Liao, C.-L. Pan, and Z. Wu, “Digital transformation and innovation and business ecosystems: A bibliometric analysis for conceptual insights and collaborative practices for ecosystem innovation,” *International Journal of Innovation Studies*, vol. 8, no. 4, pp. 406–431, Dec. 2024, doi: 10.1016/j.ijis.2024.04.003.

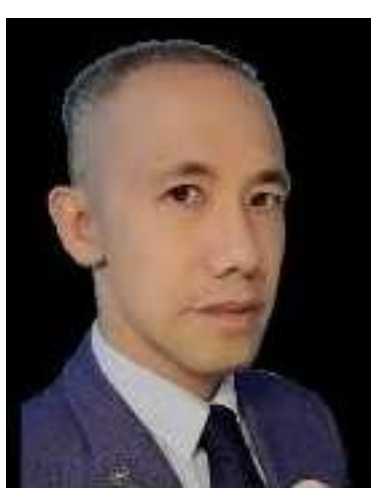

**Han-Teng** (Member, IEEE) was born in Taipei, Taiwan in 1976. He received the B.E. degree in electrical engineering (NTUEE) and the B.A. degree in Foreign Languages and Literatures from the National Taiwan University, Taipei, Taiwan, in 2000; the M.E. degree in computer science and information engineering (NTUCSIE) and the M.S.S. degree in journalism from the National Taiwan University in 2004; and the DPhil in information, communication, and the social sciences from the University of Oxford, Oxford, U.K., in 2014.

From 2004 to 2005, he was a project manager with the Open Source Software Foundry, Institute of Information Science, Academia Sinica, Taipei, Taiwan. He previously served as an Associate Professor, Vice Dean of Internet and New Media, and Director of the Higher Education Impact Assessment Center at Nanfang College of Sun Yat-Sen University, Guangzhou, China. He is currently the Project Founder of the Safe and Sustainable by Design (SSbD) Chips Value Chain Services (SCVCS) initiative and GBS.today, where he operates as an independent socio-technical strategist and data scientist based between Penang, Malaysia, and Taipei. His work integrates full-lifecycle agile product management, scientometric science mapping, and tech roadmapping to align cross-border semiconductor workflows with global regulatory frameworks.

Dr. Liao is an active member of the IEEE Technology and Engineering Management Society (TEMS) and serves on the Joint IEEE Systems Engineering/TEMS Technical Committee on Carbon Neutrality.

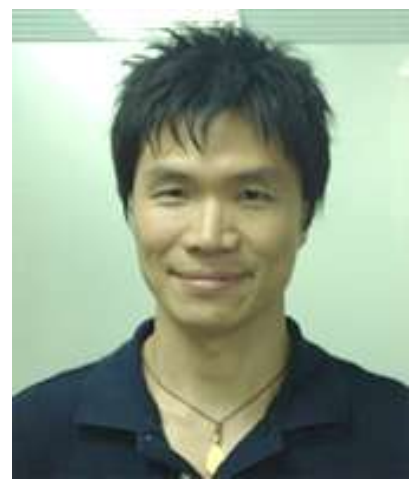

**Chang-Yi Kao** received the Ph.D. degree in computer science and information engineering from National Taiwan University of Science and Technology, Taipei, Taiwan, in 2012. He was a Senior Industry Analyst and Division Manager with the Institute for Information Industry, Taipei, Taiwan.

He is currently an Associate Professor with the Department of Computer Science and Information Management, Soochow University, Taipei, Taiwan. His research interests include technology adoption and industrial marketing, enterprise information systems design, and business intelligence.